\documentclass[onecolumn,amsmath,amsssymb,12pt]{revtex4} 
\usepackage{graphicx}
\def\lsim{\mathrel{\mathstrut\smash{\ooalign{\raise2.5pt\hbox{$<$}\cr\lower2.5pt\hbox{$\sim$}}}}}
\def\gsim{\mathrel{\mathstrut\smash{\ooalign{\raise2.5pt\hbox{$>$}\cr\lower2.5pt\hbox{$\sim$}}}}}

\def\slashed#1{{\ooalign{\hfil\hfil/\hfil\cr $#1$}}}

\begin{document}

\title{Scalar self-interactions loosen constraints from fifth force searches}

\author{Steven S. Gubser} 
\affiliation{Joseph Henry Laboratories, Princeton University, Princeton, NJ 08544, USA} 
 
\author{Justin Khoury} 
\affiliation{Institute for Strings, Cosmology and Astroparticle Physics, Columbia University, New York, NY 10027, USA} 
 
\begin{abstract} 
The mass of a scalar field mediating a fifth force is tightly constrained by experiments.  We show, however, that adding a quartic self-interaction for such a scalar makes most tests much less constraining: the non-linear equation of motion masks the coupling of the scalar to matter through the chameleon mechanism.  We discuss consequences for fifth force experiments. In particular, we find that, with quartic coupling of order unity, a gravitational strength interaction with matter is allowed by current constraints. We show that our chameleon scalar field results in experimental signatures that could be detected through modest improvements of current laboratory set-ups.
\end{abstract} 
 
\maketitle 
 
\section{Introduction}
\label{intro} 
 
We are interested in studying the dynamics of a massive scalar $\phi$ with a quartic self-interaction and weak couplings to visible matter.  The action we wish to consider is
\begin{equation}
S = \int d^4 x \, \sqrt{g} \left[ {1 \over 2} (\partial\phi)^2 - 
  {1 \over 2} m_\phi^2 \phi^2 - {\xi \over 4!} \phi^4 \right] - 
  \sum_\alpha \int_{\gamma_\alpha} ds \, m_{i_\alpha}(\phi) \,,
\label{lphi}
\end{equation}
where in the sum over $\alpha$ we give the world-line action for any number of visible particles of several species $i$, whose couplings to the scalar arises from the dependence of the masses $m_i$ on $\phi$.  Useful dimensionless measures of these couplings are 
\begin{equation}
\beta_i \equiv -M_{Pl} {d\log m_i \over d\phi} \,,
\label{BetaDef}
\end{equation}
where $M_{Pl} \equiv (8\pi G)^{-1/2}$.  The action~(\ref{lphi}) is generally covariant, but in this paper we will have occasion only to study effects in flat Minkowski space: $g_{\mu\nu} = \eta_{\mu\nu}$.

\begin{figure} 
\includegraphics[width=5in]{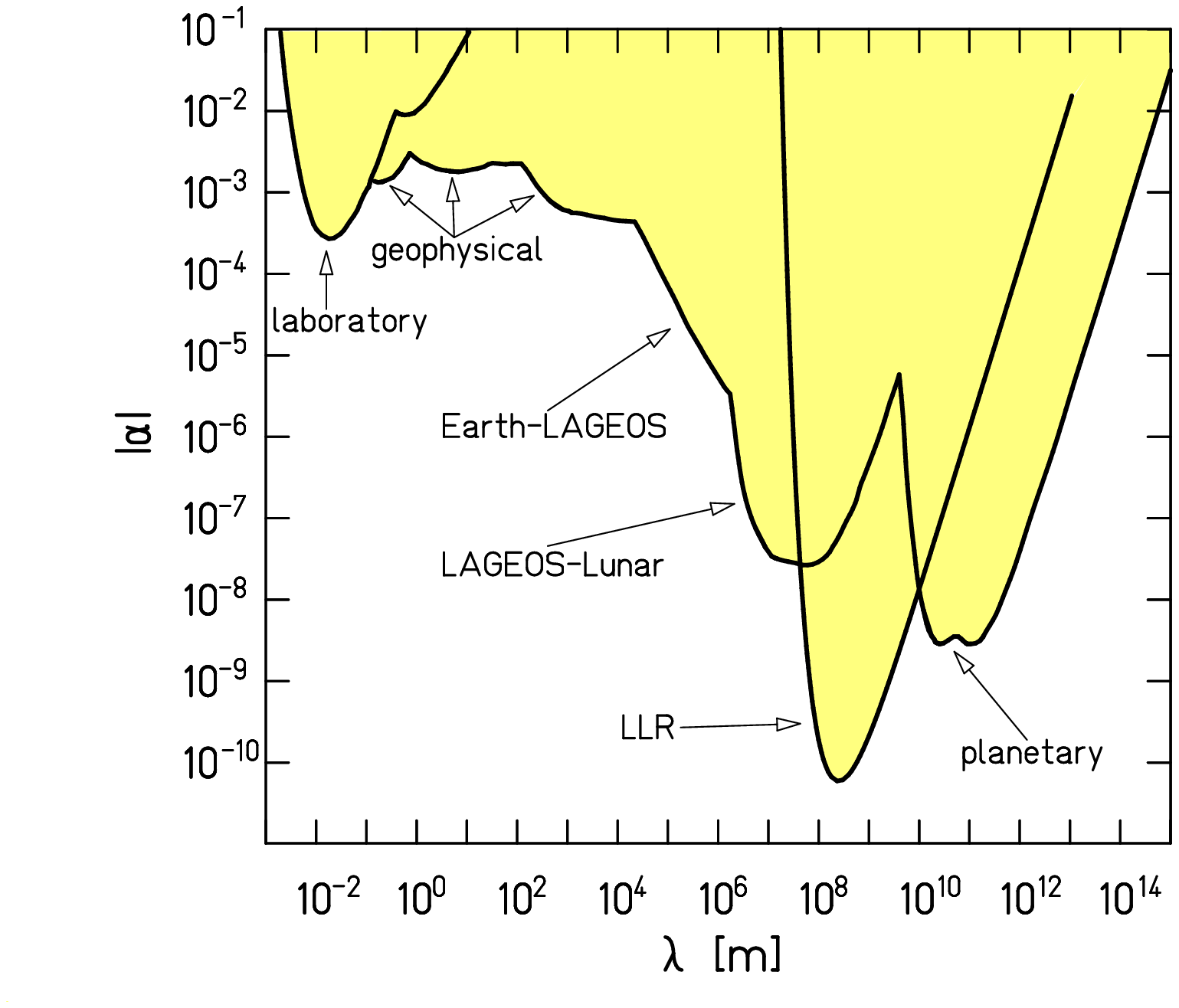}
\caption{Experimental constraints on the strength $\alpha$ and range $\lambda$(m) from all fifth force experiments to date. Note that $\alpha=1$ corresponds to gravitational strength. This plot neglects any self-coupling for the scalar field mediating the force. Reprinted from~\cite{adellong}.} 
\label{expconsa} 
\end{figure}

\begin{figure} 
\includegraphics[width=5in]{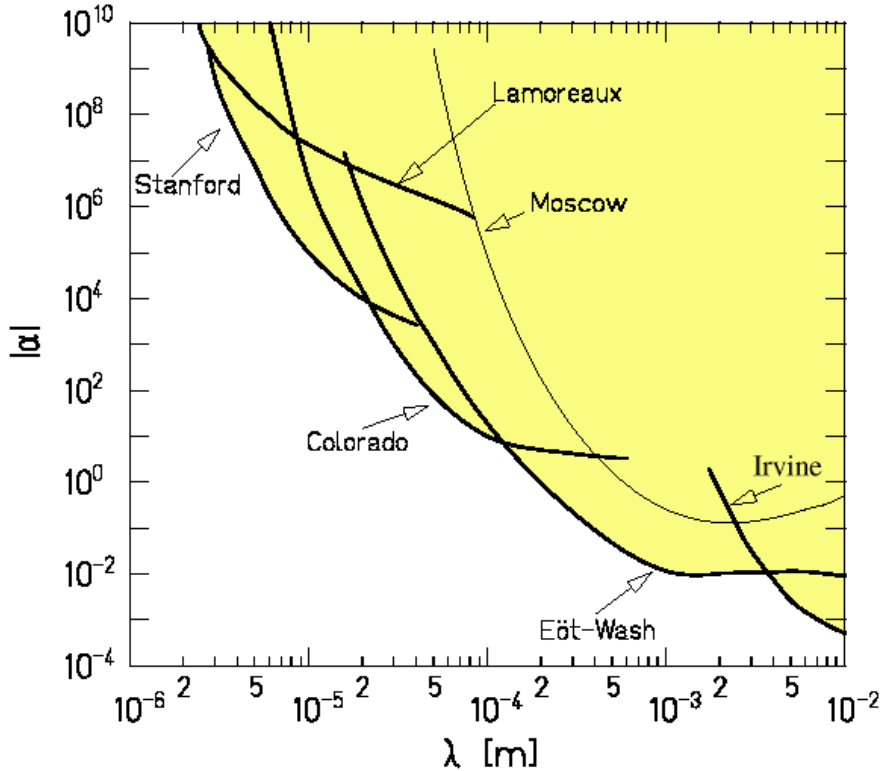}
\caption{Small-scale blow-up of previous figure. Reprinted from~\cite{adellong}.} 
\label{expconsb} 
\end{figure}

The action~(\ref{lphi}) is a standard starting point for discussions of a scalar-mediated fifth force. With $\xi=0$ and to linear order it results in a potential energy between particles of species $i$ and $j$ of the form
\begin{equation}
U(r) = -2 \beta_i \beta_j {G m_i m_j \over r} e^{-m_\phi r} \,.
\label{fifthpot}
\end{equation}
Evidently, the product $2 \beta_i \beta_j$ is the dimensionless parameter $\alpha$ referred to as the interaction strength, while $\lambda \equiv m_\phi^{-1}$ is the interaction range.  The $\beta_i$ may be different for different species of visible particles, corresponding to isotope-dependence of the scalar forces.  Let us for now ignore this possibility and take a universal value $\beta$ for all the $\beta_i$.  Experimental constraints on $\alpha$ and $\lambda$ come from a variety of sources.  The current state of affairs is shown in Figs.~\ref{expconsa} and~\ref{expconsb}~\cite{fischbach}.  In particular, we see that for $\alpha\sim {\cal O}(1)$, corresponding to gravitational strength, there is no evidence for a fifth force over the range $100\;\mu{\rm m} \lsim \lambda \lsim 10^4$~AU.

\begin{figure} 
\includegraphics[width=5in]{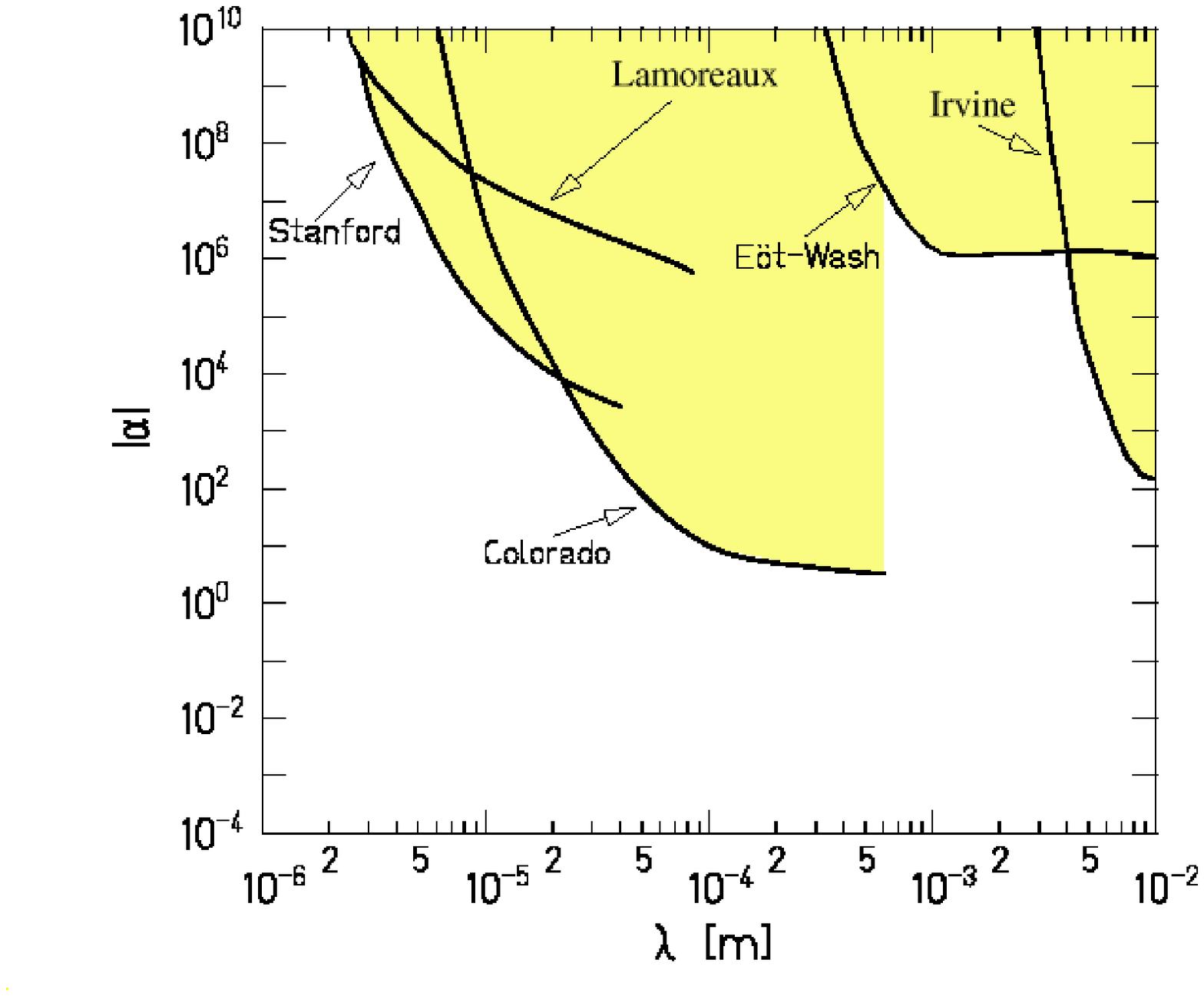}
\caption{Revised experimental constraints on the strength $\alpha$ and range $\lambda$(m) including the effects of a quartic self-interaction term with $\xi=1$. The main difference is for the E$\ddot{{\rm o}}$t-Wash~\cite{adel} and Irvine~\cite{irvine} experiments since their test masses have a thin shell. The test masses used in the Stanford~\cite{stan} and Colorado~\cite{col} experiments, on the other hand, are too small to have a thin shell. Thus the corresponding curves are the same as in Fig.~\ref{expconsb}.}
\label{expconsbrevised} 
\end{figure}

If we set $\xi=1$ in Eq.~(\ref{lphi}), then the experimental constraints on $\alpha$ and $\lambda$ are drastically altered through the chameleon mechanism~\cite{cham}.  In particular, we find that the bounds from geophysical, LAGEOS satellite, lunar laser ranging (LLR), and planetary data become essentially irrelevant. For instance, the new LAGEOS bound is $\alpha < 10^{36}$. The revised constraints from laboratory experiments are shown in Fig.~\ref{expconsbrevised}. 
Our main result is that $\alpha\sim 1$ is allowed on all scales with a quartic self-interaction term.

For any given $m_\phi$, it would seem that to allow $\xi$ to be of order unity in Eq.~(\ref{lphi}) is a more generic circumstance than constraining it to be negligibly small. Paradoxically, it is {\it less} generic from the point of view of naturalness in quantum field theory: more precisely, Eq.~(\ref{lphi}) with $\xi \sim {\cal O}(1)$ and $1/m_\phi \gsim 100\;\mu{\rm m}$ is fine-tuned in the presence of otherwise natural couplings to the fields of the Standard Model.  We will give a more detailed analysis of this in section~\ref{NATURALNESS}.  A failure of naturalness does amount to a serious obstacle to accepting the starting point action~(\ref{lphi}) as theoretically well-motivated.  However, we take the view that the predictions following from it are sufficiently distinctive and testable that it is worth laying them out as an alternative to results based on making $\phi$ a free field.  Also, in section~\ref{NATURALNESS}, we explain how the obstacles to naturalness can be to a great extent overcome by positing a coupling of $\phi$ to a component of an isospin vector. 

In section~\ref{CHAMELEON} we describe the chameleon mechanism as it applies to the model~(\ref{lphi}) and point out that with $\xi \sim {\cal O}(1)$, $\beta \sim {\cal O}(1)$, and $m_\phi=0$, the effective range of the scalar force depends strongly on the density of the ambient medium, being roughly a millimeter in Earth's atmosphere and roughly $10\,{\rm km}$ at a generic point in the solar system.  This is one of the key points which relaxes experimental constraints on $\alpha$ and $\lambda$ when a quartic coupling is turned on.  Another key point is the failure of the superposition principle and the shell theorem that inevitably accompanies the non-linear equation of motion for $\phi$.  A useful intuition is that for most objects, only a thin shell around the edge of the object exerts a significant ``pull'' on $\phi$, further suppressing the strength of the $\phi$-mediated force.  In section~\ref{revthin} we will describe this in more quantitative detail.  Then in section~\ref{mod5th} we will describe the modified constraints on $\alpha$ and $\lambda$ with $\xi=1$. We summarize our results and discuss prospects for future experiments in section~\ref{conclu}.

\section{The chameleon mechanism}
\label{CHAMELEON}

Briefly, the chameleon mechanism is the generation of an effective mass for a scalar through the interplay of scalar self-interactions and interactions with ambient matter.  Let us again assume a universal value $\beta$ for the dimensionless couplings $\beta_i$, and let us treat visible sector matter as a perfect non-relativistic fluid with negligible pressure.  Over-simplifying in a way that will not affect the final results, we replace visible sector matter by a single massive species with number density $n$ and particle mass $m(\phi)$.  From Eq.~(\ref{lphi}) and a hydrodynamic approximation one can obtain the following equation of motion for $\phi$:
\begin{eqnarray}
-\nabla^\mu \partial_\mu \phi &=& m_\phi^2 \phi + 
  {\xi \over 3!} \phi^3 - n {dm \over d\phi} 
 = {dV_{eff} \over d\phi} \label{eom} \\
V_{eff} &\equiv& {1 \over 2} m_\phi^2 \phi^2 + {\xi \over 4!} \phi^4 + 
  n m(\phi) \nonumber \\
  &\approx&
  {1 \over 2} m_\phi^2 \phi^2 + {\xi \over 4!} \phi^4 - 
  {\beta\phi \over M_{Pl}} \rho + (\phi\hbox{-independent}) \label{veff} \,,
\end{eqnarray}
where in the last, approximate equality we have used Eq.~(\ref{BetaDef}), expanded $m(\phi)$ to linear order in $\phi$, and identified $nm(\phi)$ as the energy density $\rho$.  We assume that $m(\phi)$ is a smooth enough function for the linear approximation to be valid for the following discussion.

Evidently, the equilibrium value of $\phi$ in the presence of non-vanishing $\rho$ is not $0$, but rather the minimum $\phi_{min}$ of $V_{eff}(\phi)$.  Small fluctuations around this equilibrium obey a Klein-Gordon equation with a mass given by
\begin{equation} 
m^2_{eff} \equiv \frac{d^2V_{eff}(\phi_{min})}{d\phi^2} = m_\phi^2 + \frac{\xi}{2}\phi_{min}^2 + n {d^2 m(\phi_{min}) \over d\phi^2} \,, 
\label{mmin} 
\end{equation} 
where the last term would be neglected in the linear approximation for $m(\phi)$.  Clearly, the effective mass is larger for larger $\rho$: the scalar force is screened, because of the quartic self-interaction, in the presence of matter.

This chameleon mechanism is most spectacular when $m_\phi=0$.  Then, using the linear approximation for $m(\phi)$,
\begin{eqnarray} 
\phi_{min} &=& \left(\frac{6\beta\rho}{\xi M_{Pl}^4}\right)^{1/3}M_{Pl} \label{ptemp} \\ 
m^2_{eff} &=& \frac{\xi}{2}\phi_{min}^2 = \left(\frac{9\xi\beta^2}{2}\right)^{1/3}\left(\frac{\rho}{M_{Pl}^4}\right)^{2/3}M_{Pl}^2\,.
\label{mmin0} 
\end{eqnarray}
In more user-friendly units:
\begin{eqnarray} 
\phi_{min} [{\rm mm}^{-1}] &\approx& 10\left(\frac{\beta}{\xi}\right)^{1/3}(\rho[{\rm g}/{\rm cm}^3])^{1/3}\, \label{pfriendly} \\
m_{eff}^{-1}[{\rm mm}] &\approx& 0.1\;(\xi\beta^2)^{-1/6}(\rho[{\rm g}/{\rm cm}^3])^{-1/3}\,. 
\label{friendly} 
\end{eqnarray} 
Corrections to these formulas would be suppressed by factors of $\rho/M_{Pl}^4$ or $\phi_{min}\ll M_{Pl}$, both of which are indeed small quantities for any reasonable matter density. 
 
For example, the atmosphere has mean density $\rho\approx 10^{-3}\;{\rm g}/{\rm cm}^3$. Substituting in Eq.~(\ref{friendly}) and assuming $\beta,\xi\sim {\cal O}(1)$, we find 
\begin{equation} 
m_{atm}^{-1} \approx 1\;{\rm mm}\,. 
\label{matm} 
\end{equation} 
Hence, even though the field is exactly massless in vacuum, we see that the force it mediates has a range of one millimeter in the atmosphere. 
 
In the solar system, the relevant matter background is the nearly homogeneous baryonic gas and dark matter, with density $\rho_G\approx 10^{-24}\;{\rm g}/{\rm cm}^3$. The corresponding interaction range is then 
\begin{equation} 
m_G^{-1} \approx 10\;{\rm km}\,. 
\label{mG} 
\end{equation} 
 
In~\cite{cham}, the chameleon mechanism relied upon the interplay of a runaway potential for $\phi$ and exponential couplings to matter, and one scale had to be set by hand in order to achieve something similar to Eq.~(\ref{matm}).  Thus, a novelty of the present setup is that millimeter range screening comes to us ``for free'' given a quartic self-interaction with a coefficient of order unity.  Another distinction in comparing with~\cite{cham} is that the $\phi$-mediated interactions are short range both in the atmosphere and in the solar system.

\section{Naturalness of the $\phi^4$ chameleon construction}
\label{NATURALNESS}

Consider the following alternative to our starting point action~(\ref{lphi}):
\begin{equation}
S = \int d^4 x \, \left[ {1 \over 2} (\partial\phi)^2 - 
 {1 \over 2} m_\phi^2 \phi^2 - {\xi \over 4!} \phi^4 + 
 \bar\psi i \slashed\partial \psi - m(\phi) \bar\psi\psi \right] \,,
\label{DiracAction}
\end{equation}
where $\psi$ is a Dirac fermion.  Once again, we have over-simplified considerably by replacing visible sector matter with a single free fermion, but the action~(\ref{DiracAction}) will nevertheless serve to illustrate points that carry over to a more general setting.  At the classical level, for non-relativistic processes at low enough density that the Pauli Exclusion Principle is not important, Eq.~(\ref{DiracAction}) is equivalent to Eq.~(\ref{lphi}).  It is assumed that $m(\phi) = m(0) h(\beta\phi/M_{Pl})$ for some function $h(x) = 1-x+{\cal O}(x^2)$, whose further Taylor coefficients are at most of order unity.

\begin{figure} 
\includegraphics[width=5in]{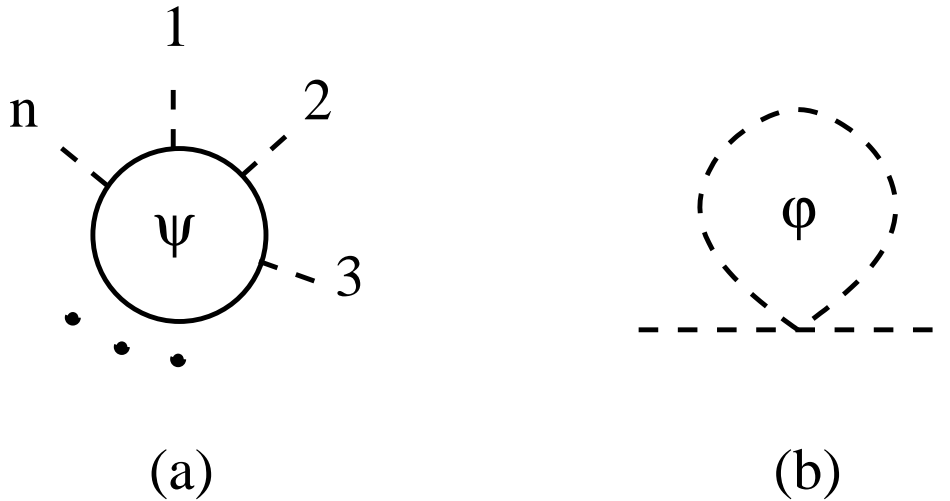}
\caption{Loop diagrams for the action~(\ref{DiracAction}) which contribute to the scalar potential.} 
\label{loops}
\end{figure}

If $\xi=0=m_\phi$, then integrating out $\psi$ via graphs of the form shown in Fig.~\ref{loops}a leads to a potential for $\phi$ of the general form $V_{eff}(\phi) = \Lambda^4 g(\beta\phi/M_{Pl})$ for some function $g(x)$ whose values and derivatives are of order unity (apart from logarithmic singularities at points where $m(\phi)=0$).  A local minimum of $V_{\rm eff}(\phi)$ would typically give a mass for $\phi$ of the form $m_{eff}^2 \sim \Lambda^4 \beta^2/M_{Pl}^2$.  If $\Lambda \sim 5\;{\rm TeV}$---a reasonable scale for present purposes since we have no direct knowledge of physics above the TeV scale---then $m_{eff}[{\rm meV}] \sim 10\beta$.  (The coefficient is slightly lower if one accounts for phase space factors \cite{adellong}.)  This is a standard line of argument, whose conclusion is that a range on the order of a millimeter for a scalar-mediated force is natural given a cutoff near the electroweak scale and Planck-suppressed couplings.

Modern-day experiments probe a much broader range of lengths, despite the argument just summarized that naturalness favors the sub-millimeter regime.  Let us then regard naturalness as a guide to favored regimes but not as a firm rule.  In this spirit, it is instructive to see how naturalness becomes more difficult to arrange in a theory with $\xi$ of order unity.  At the end of this section, we will suggest a way to circumvent the main difficulty.

As a warmup, let's first consider the case where $\xi \sim 1/10$ but the fermions are decoupled ({\it i.e.},~$dm/d\phi=0$).  Then having small or zero $m_\phi$ is no less natural  than when $\xi$ is $0$ \footnote{Actually, this is only true once a symmetry breaking issue is resolved, as we will soon describe.}.  This seems to conflict with the prediction that the self-energy graph in Fig.~\ref{loops}b contributes $\xi\Lambda^2$ to $m_\phi^2$.  Let us therefore analyze the issue carefully in a renormalization group (RG) language.

Assume then that $\xi = \xi(\Lambda) = 1/10$ and $m_\phi = 0$ in the action that best describes the dynamics at a scale $\Lambda$ (this is an effective action in the Wilsonian sense, and the specific value of $\xi(\Lambda)$ is chosen so as to have some control in perturbation theory).  Working to one-loop order, the effective potential (part of the 1PI effective action) is (see for instance \cite{ZeeBook})
\begin{equation}
V_{eff}(\phi) = {\xi(\mu) \over 4!} \phi^4 + 
 {\xi(\mu)^2 \over (16\pi)^2} \phi^4
  \left( \log {\phi^2 \over \mu^2} - {25 \over 6} \right) \,,
\label{Veff}\end{equation}
where the renormalization conditions are ${d^2 V \over d\phi^2}(0) = 0$ and ${d^4 V \over d\phi^4}(\mu) = \xi(\mu)$, and the fact that $V_{eff}$ is independent of the arbitrary scale $\mu$ is encoded in the RG equation $\mu {d\over d\mu} \xi = {3 \over 16\pi^2} \xi^2$.

A difficulty with~(\ref{Veff}) is that the $\phi^4 \log\phi^2$ term causes symmetry breaking: the minimum is at some non-zero $\phi$, and $\phi$ at this minimum has a mass.  This is easily cured by adding a fermion, call it $\chi$, with mass $m_\chi(\phi) = \eta \phi$, where $\eta$ is some additional coupling of order unity.  This fermion is {\it not} part of the Standard Model: it is an extra field whose sole purpose is to reverse the sign on the $\phi^4 \log\phi^2$ term to insure a stable minimum at $\phi=0$.  The main point, which the fermions do not alter, is that $m_\phi$ is absent at lower scales if it is absent at a high scale.  If $m_\phi \neq 0$, then it evolves according to an RG equation:
\begin{equation}
\mu {d\over d\mu} m_\phi^2 = \left( {\xi \over 16\pi^2} +
 {\eta^2 \over 4\pi^2} \right) m_\phi^2 \,.
\label{PhiEvolution}\end{equation}
This multiplicative renormalization of $m_\phi$ may seem in conflict with the idea that the self-energy graph in Fig.~\ref{loops}b tends to ``drag'' $m_\phi$ up to the cutoff $\Lambda$.  The resolution is that adding $\xi \Lambda^2$ to $m_\phi^2$ is (part of) how one gets from a classical action to a quantum effective action, whereas the renormalization equation~(\ref{PhiEvolution}) is how one gets from a Wilsonian action for the quantum theory at one scale to the effective action at a lower scale.

From the Wilsonian perspective, then, before adding fermions, non-zero $\xi$ with $m_\phi=0$ is indeed more generic than $\xi=0=m_\phi$.  Supersymmetry and / or string theory however provide numerous examples of supersymmetric field theories with flat directions or string moduli, for which $\xi=0=m_\phi$ may indeed be nearly right at some high energy scale.  How the concept of genericity may be altered in string theory is only beginning to be understood.  

There are examples in supersymmetric field theory where the renormalizable scalar potential is purely quartic: for example, consider a $U(1)$ gauge theory coupled to three chiral superfields, one of charge $1$ and the others of charge $-1/2$.  Gauge invariance restricts the superpotential to purely cubic terms.  The potential receives quartic contributions from D-terms and F-terms, and cases with a unique global minimum at the origin are readily constructed.  It is more difficult to couple such a theory to the MSSM in such a way that the masses of quarks $m_q(\phi)$ give $\beta = -M_{Pl} d\log m/d\phi$ on the order of unity, where $\phi$ denotes the scalars charged under the extra $U(1)$.

Let us put supersymmetry aside and simply assume that the scalar potential is ${\xi \over 4!} \phi^4$ at some high scale $\Lambda$.  The real problem arises when we include loops of Standard Model fields, because the latter tend to change the potential in such a way as to shift the vacuum away from $\phi=0$.  To see this, it is enough to go back to the action~(\ref{DiracAction}) and compute the one-loop effective potential: before any renormalization,
\begin{equation}
V_{eff}(\phi) = V_{bare}(\phi) + 
 {\Lambda^2 \over 32\pi^2} \sum_i (-1)^{F_i} m_i(\phi)^2 - 
 {1 \over 64\pi^2} \sum_i (-1)^{F_i} m_i(\phi)^4
  \log {\sqrt{e} \Lambda^2 \over m_i(\phi)^2} \,,
\label{SeveralSpecies}
\end{equation}
where the sum is over the degrees of freedom (counting a real boson as $1$ and a complex Dirac fermion as $4$), and $(-1)^F$ is $1$ for a boson and $-1$ for a fermion.  The quantity $m_\phi(\phi)$ appearing in~(\ref{SeveralSpecies}) is by definition $d^2 V_{eff}(\phi)/d\phi^2$.  Evidently, $V_{eff}(\phi)$ has odd terms in $\phi$ even if $V_{bare}(\phi)$ doesn't.  So $\phi=0$ isn't even an extremum of $V_{eff}(\phi)$, and at the true extremum $\phi$ will have a mass as well as cubic and and higher interactions.  We do not see how to make an interesting theory of chameleon fields in this way.

The problem originates in the loss of the ${\bf Z}_2$ symmetry $\phi \to -\phi$ that arises when we couple to a fermion with mass $m(\phi) = m_0 (1+\beta\phi/M_{Pl})$.  A chiral phase rotation $\psi \to e^{i\pi\gamma_5} \psi$ can switch the sign of $m_0$ but not of $\beta$.  (This explains why the fermion $\chi$ mentioned earlier, with $m_\chi(\phi) = \eta\phi$, {\it doesn't} break the ${\bf Z}_2$: one just extends its action to include $\chi \to e^{i\pi\gamma_5} \chi$.)  Let us try to get closer to a field-theoretically natural chameleon by considering $\phi$ coupled to up and down quarks through a term that is part of an isospin triplet:
\begin{equation}
S = \int d^4 x \sqrt{g} \, \left( {1 \over 2} (\partial\phi)^2 - 
 {1 \over 2} m_\phi^2 \phi^2 - {\xi \over 4!} \phi^4 + 
 \bar\chi (i\slashed\partial - \eta\phi) \chi + 
 \bar q (i\slashed\partial - m) q - 
 \tilde\beta \phi \bar q \tau_3 q \right) \,,
\label{BigAction}\end{equation}
where $q$ denotes the isospin quark doublet $\left( {u \atop d} \right)$, and $\tau_3 = \left( 1\ \ 0 \atop 0\ -1 \right)$ is one of the Pauli matrices.  The action~(\ref{BigAction}) is still conspicuously imperfect: the quarks are free except for the interaction with $\phi$, and we have given them an equal mass $m$.  But the $\phi \to -\phi$ symmetry can now be extended to include an isospin rotation sending $u \to d$ and $d \to -u$, and the effective potential, for $\xi/\eta^2$ not too large, is even with a minimum at $\phi=0$.

In a more realistic treatment including QCD, there is chiral symmetry breaking through an isospin-singlet VEV $\langle \bar q q \rangle$, but $\langle \bar q \vec{\tau} q \rangle = 0$ to avoid breaking isospin.  The absence of an isospin triplet VEV is fortunate because otherwise there would be a tadpole for $\phi$ that would lead us back to $V_{eff}(\phi)$ that is not even.

If isospin were a perfect symmetry of QCD, then the coupling $-\tilde\beta \phi \bar q \tau_3 q$ would result in a naturally even scalar potential with small $m_\phi$ preserved by RG flow, and opposite dimensionless couplings $\beta_i$ (defined as in~(\ref{BetaDef})) for the proton and neutron, proportional to $\tilde\beta$.  Isospin is slightly broken by explicit quark masses and electromagnetic effects, and we defer to the future an attempt to determine how much this alters the final picture.

In summary, the simplest model that we study, where $m_\phi=0$ and the $\beta_i$ for visible particles are all equal to a common value, is not field-theoretically natural because of terms in $V_{eff}(\phi)$ that violate $\phi \to -\phi$ symmetry.  One could contrive to eliminate such terms by cancelling contributions to $V_{eff}(\phi)$ from the first generation of quarks against contributions from the other generations: this is obviously a fine-tuned approach since the interactions of the heavy quarks are quite different from $u$ and $d$.  An alternative model, with equal and opposite $\beta_i$ for protons and neutrons, is more promising from the perspective of naturalness.

\section{Review of the thin-shell mechanism} \label{revthin} 
 
In the previous sections we have encountered one of the characteristic features of chameleons, namely that their mass depends on the surrounding density. Another important property of such scalars is that their effective coupling to matter is also density-dependent. To be precise, the $\phi$-field outside a body involves an effective coupling, $\beta_{eff}$, which, for sufficiently dense objects, is much smaller than the ``bare'' coupling $\beta$. Thus the $\phi$-force between two dense bodies is suppressed by this effect, first referred to as a ``thin-shell'' mechanism in~\cite{cham} for reasons that will soon be obvious. Density-dependent effective couplings between a scalar field and matter were first observed by Damour and Esposito-Far\`ese~\cite{gef}. 
 
Consider a spherical body of mass $M_c$, radius $R_c$ and homogeneous density $\rho_c$. The body is assumed immersed in a homogeneous medium of density $\rho_\infty$. Throughout we denote by $\phi_c$ and $\phi_\infty$ the field value that minimizes $V_{eff}$ for $\rho=\rho_c$ and $\rho_\infty$, respectively. Similarly, the mass of small fluctuations about these minima are $m_c$ and $m_\infty$. Equation~(\ref{eom}) thus reduces to 
\begin{equation} 
\frac{d^2\phi}{dr^2} + \frac{2}{r}\frac{d\phi}{dr} = m^2_\phi\phi+\frac{\xi}{3!}\phi^3 - \frac{\beta}{M_{Pl}}\rho(r)\,, 
\label{eomthin} 
\end{equation} 
where $\rho=\rho_c$ for $r<R_c$ and $=\rho_\infty$ for $r>R_c$. This differential equation is subject to the following boundary conditions: $i)$ the solution must be regular at the origin; $ii)$ far away from the body, the field must tend to its expectation value in the ambient medium. The latter ensures that the force on a test particle becomes vanishingly small as the test particle moves infinitely far. Thus, we impose
\begin{eqnarray} 
\nonumber 
& & \frac{d\phi}{dr} = 0 \qquad \;{\rm at}\;\;\;r=0\,; \\ 
& & \phi\rightarrow \phi_\infty \qquad {\rm as}\;\;\; r\rightarrow \infty\,. 
\label{bc} 
\end{eqnarray} 

Although the above equation is non-linear, we can use the linear approximation which, as we will see, is valid for all objects of interest. For $r<R_c$, substituting $\phi=\phi_c + \delta\phi$ in Eq.~(\ref{eomthin}) gives
\begin{equation}
\frac{d^2\delta\phi}{dr^2} + \frac{2}{r}\frac{d\delta\phi}{dr} = m_c^2\delta\phi\,.
\label{int1}
\end{equation}
The solution satisfying the first boundary condition above is
\begin{equation}
\phi(r<R_c) = \frac{A\sinh{m_cr}}{r} + \phi_c\,,
\label{inside}
\end{equation}
where $A$ is a constant to be determined later. 

Similarly, far away from the body, we can once again linearize Eq.~(\ref{eomthin}), this time about $\phi_\infty$. The solution satisfying $\phi\rightarrow\phi_\infty$ as $r\rightarrow\infty$ is
\begin{equation}
\phi(r\gg R_c) = \frac{Be^{-m_\infty r}}{r}+\phi_\infty\,,
\label{asymp}
\end{equation}
where $B$ is another constant. 

Strictly speaking, this solution is only valid in the asymptotic region far from the object. Nevertheless, to obtain an approximate analytical solution, we will assume that it holds even near the surface, $r\sim R_c$. We will later determine the degree of accuracy of this approximation by comparing with numerical calculations. Imposing continuity of $\phi$ and $d\phi/dr$ at $r=R_c$ then allows one to solve for $A$ and $B$. This gives
\begin{equation}
\phi(r<R_c) =  \left\{\frac{1+ m_\infty R_c}{m_cR_c\coth(m_cR_c)+m_\infty R_c}\right\} \frac{R_c(\phi_c-\phi_\infty)}{\sinh(m_cR_c)}\frac{\sinh(m_cr)}{r} + \phi_c\,;
\label{linsolin}
\end{equation}
\begin{equation}
\phi(r>R_c) =  \left\{\frac{m_cR_c\coth(m_cR_c)-1}{m_cR_c\coth(m_cR_c)+m_\infty R_c}\right\} \frac{R_c(\phi_c-\phi_\infty)}{r}e^{-m_\infty(r-R_c)} + \phi_\infty\,.
\label{linsolout}
\end{equation}

This, however, assumes linearity for $r<R_c$. It is easily seen from Eq.~(\ref{linsolin}) that this is a valid assumption provided that
\begin{equation}
m_cR_c\gg 1\,.
\label{cond}
\end{equation}
If we further assume that the density contrast between the object and the ambient matter is high, so that $\phi_c\gg\phi_\infty$ and $m_c\gg m_\infty$, the exterior solution in Eq.~(\ref{linsolout}) reduces to
\begin{equation}
\phi(r>R_c) \approx \frac{\beta_{eff}}{4\pi M_{Pl}}\frac{M_ce^{-m_\infty(r-R_c)}}{r} + \phi_\infty\,,
\label{thinsoln}
\end{equation}
with
\begin{equation}
\beta_{eff}=\frac{3\phi_c M_{Pl}}{\rho_c R_c^2} = \left(\frac{3}{m_c R_c}\right)^2\beta\,, 
\label{betaeff} 
\end{equation} 
where in the last step we have used Eqs.~(\ref{ptemp}) and~(\ref{mmin0}). Equation~(\ref{thinsoln}) is precisely the exterior solution for a normal scalar of mass $m_\infty$, with $\beta_{eff}$ replaced by $\beta$. Thus, due to the quartic self-coupling, the exterior solution for objects satisfying $m_c R_c\gg 1$ is exactly that a point particle albeit with a much weaker effective coupling constant, $\beta_{eff}\ll\beta$. The physical interpretation is clear: non-linear interactions responsible for this suppressed coupling are only effective provided the object is sufficiently large compared to the Compton wavelength of the chameleon, so that the latter can essentially ``feel'' the object.

This suppression mechanism was called ``thin-shell'' in~\cite{cham}. This terminology is motivated by the realization from Eq.~(\ref{inside}) that $\phi$ stays essentially constant throughout the bulk of the object except within a thin shell of thickness $\sim m_c^{-1}$. Thus only this thin shell contributes to the force on a test particle, resulting in a suppressed effective coupling $\beta_{eff}$.

To corroborate these statements, consider the opposite limit, $m_cR_c\ll 1$. In this case, the linear approximation does not hold, and therefore $\phi\ll\phi_c$ for $r<R_c$. Equation~(\ref{eomthin}) in the range $0<r<R_c$ thus becomes
\begin{equation} 
\frac{d^2\phi}{dr^2} + \frac{2}{r}\frac{d\phi}{dr} \approx -\frac{\beta}{M_{Pl}}\rho_c\,, 
\end{equation} 
with solution 
\begin{equation} 
\phi(r<R_c)=-\frac{\beta\rho_cr^2}{6 M_{Pl}} + const.
\label{inside2}
\end{equation} 
The exterior solution ($r>R_c$) is once again of the form $\phi(r)\sim e^{-m_\infty r}/r$. Matching $\phi$ and $d\phi/dr$ at the surface of the object gives 
\begin{equation} 
\phi(r>R_c) \approx \frac{\beta}{4\pi M_{Pl}}\frac{M_ce^{-m_\infty(r-R_c)}}{r} + \phi_\infty\,.
\label{thicksoln} 
\end{equation} 
As anticipated, the exterior field is that of a point particle with unsuppressed coupling in this case. Moreover, there is no thin shell since, as seen from Eq.~(\ref{inside2}), the field never remains nearly constant within the object.

\begin{figure} 
\includegraphics[width=3in]{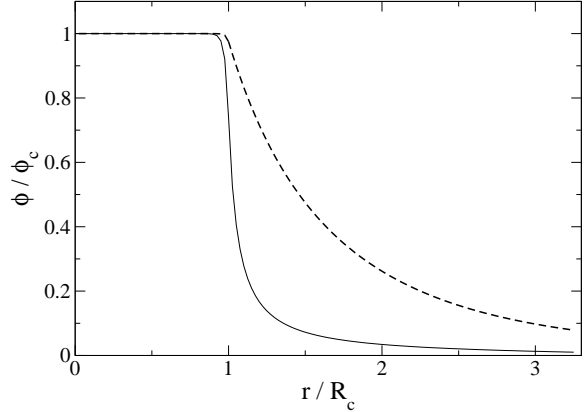}
\caption{Profile for the chameleon for an object with $m_cR_c\approx 65$. The solid curve is the result of numerical integration; the dotted line is the analytical approximation derived in the text.} 
\label{phi4yes} 
\end{figure}

The above analytic expressions have been checked against numerical calculations. Consider $\beta=1$, $\xi=1$, $\rho_c=1\;{\rm g}/{\rm cm}^3$ and $\rho_\infty=10^{-6}\;{\rm g}/{\rm cm}^3$, corresponding roughly to a ball of Beryllium in thin air. Correspondingly, from Eqs.~(\ref{friendly}) we have $\phi_c\approx m_c\approx 10\;{\rm mm}^{-1}$ and $\phi_\infty\approx m_\infty \approx 0.1\;{\rm mm}^{-1}$. Figure~\ref{phi4yes} shows the result of numerically integrating Eq.~(\ref{eomthin}) for an object of radius $R_c\approx 6.5\;{\rm mm}$. Since $m_cR_c\approx 65$, we expect a suppressed coupling in this case. This is confirmed by the numerics. The solid curve in Fig.~\ref{phi4yes} is the numerical solution. We see that indeed the field remains at $\phi\approx\phi_c$ for $0<r\lsim R_c$, justifying the linear approximation in this regime and supporting the ``thin-shell'' interpretation. The dotted line is a plot of Eqs.~(\ref{linsolin}) and~(\ref{linsolout}). For $r>R_c$, the exact solution is initially steeper than the analytic approximation; very quickly, however, the roles are reversed, and it is the dotted line that it is steeper. The two curves eventually both decay exponentially. 

\begin{figure} 
\includegraphics[width=3in]{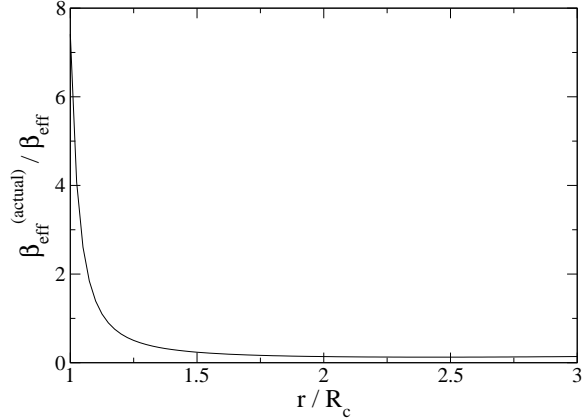}
\caption{Ratio of the numerical to theoretical chameleon-mediated force on a test particle in the exterior of the object.}
\label{comp} 
\end{figure}

The small discrepancy for $r\gsim R_c$ is a consequence of our trusting Eq.~(\ref{asymp}) near the surface of the object. To assess how this affects our predictions for the suppressed coupling, let us define
\begin{equation}
\beta_{eff}^{(actual)}\equiv \beta_{eff}\frac{\nabla\phi^{(actual)}}{\nabla\phi^{(theo)}}\,,
\end{equation}
where the superscripts $(actual)$ and $(theo)$ refer to the numerical and analytical solutions, respectively, and where $\beta_{eff}$ is given in Eq.~(\ref{betaeff}). Thus $\beta_{eff}^{(actual)}/\beta_{eff}$ is the ratio of the numerical to theoretical $\phi$-mediated force on a test particle. Fig.~\ref{comp} is a  plot of this ratio. We see that the discrepancy ranges from $\approx 7$ at $r=R_c$ to $\approx 0.1$ for $r\gg R_c$. In other words, the suppressed coupling is found numerically to lie in the range $2\cdot 10^{-4}\lsim\beta_{eff}^{(actual)}\lsim 1.5\cdot 10^{-2}$, while the analytical estimate in Eq.~(\ref{betaeff}) gives $\beta_{eff}\approx 2\cdot 10^{-3}$. The discrepancy disappears, of course, as $m_cR_c\rightarrow 1$. In particular, the force between two macroscopic test masses is always less than or equal to the force between two point particles of the same mass.

\section{Modified fifth force constraints} \label{mod5th} 

We have seen that a quartic interaction suppresses the $\phi$-mediated force in two ways. Firstly, the ambient matter density gives a non-zero expectation value to $\phi$ about which the mass of fluctuations can be much larger than in vacuum. Secondly, non-linear interactions of $\phi$ inside a dense macroscopic body leads to a suppressed effective coupling as felt by an outside test mass. In this section, we explore the consequences of these two effects for fifth force searches.

Consider two bodies of radius $R_i$ and mass $M_i$, $i=1,2$, both satisfying Eq.~(\ref{cond}). For simplicity, we assume that they have equal bare coupling $\beta$ and density $\rho=1\;{\rm g}/{\rm cm}^3$. The latter is the ballpark value for all bodies used for fifth force constraints, let it be test masses in the laboratory, satellites or planets. From the second of Eqs.~(\ref{friendly}), this corresponds to $m_c^{-1}\approx 0.1\;(\xi\beta^2)^{-1/6}$~mm. The potential energy associated with the fifth force is then
\begin{equation}
U(r) = -\alpha_{eff} \frac{G M_i M_j}{r} e^{-r/\lambda_{eff}} \,,
\label{fifthpoteff}
\end{equation}
where $\alpha_{eff}$ is given by Eq.~(\ref{betaeff}) and the relation $\alpha=2\beta^2$:
\begin{equation}
\alpha_{eff} \sim 10^{-2}\left(\frac{\alpha}{\xi^2}\right)^{1/3} \frac{1}{R_1^2[{\rm mm}]R_2^2[{\rm mm}]}\,, 
\label{betaeffbis} 
\end{equation} 
and where $\lambda_{eff}$ is the effective Compton wavelength of $\phi$ in the surrounding medium. From Eqs.~(\ref{mmin0}) and~(\ref{friendly}), we find for the latter 
\begin{equation}
\lambda_{eff} = \frac{\lambda}{\sqrt{1+100(\xi\beta^2)^{1/3}\rho_\infty^{2/3}[{\rm g}/{\rm cm}^3]\lambda^2[{\rm mm}]}}\,.
\label{lambdaeff}
\end{equation}

\subsection{Laboratory Searches}
\label{LabSearches}

Laboratory fifth force experiments are performed in vacuum, $\rho_\infty\approx 0$, corresponding to an unsuppressed interaction range:
\begin{equation}
\lambda_{eff}=\lambda\,.
\end{equation}
The tightest constraints from the laboratory on forces mediated by scalars with insignificant self-interaction come from E$\ddot{{\rm o}}$t-Wash~\cite{adel} and Irvine~\cite{irvine}. Typical test masses for these experiments have characteristic size $R_i\sim 10$~mm, resulting in
\begin{equation}
\alpha_{eff} \sim 10^{-6}\left(\frac{\alpha}{\xi^2}\right)^{1/3}\,.
\label{alpefflab}
\end{equation}
To be completely accurate, however, in laboratory experiments the separation between test masses is often smaller than their size. We have seen in Sec.~\ref{revthin} that, in this regime, the prediction for $\beta_{eff}$ is smaller by a factor of 10 or so compared to the numerical solution. Including this fudge factor, we obtain
\begin{equation}
\alpha_{eff} \sim 10^{-4}\left(\frac{\alpha}{\xi^2}\right)^{1/3}\,.
\label{alpefflab2}
\end{equation}
It is the effective coupling $\alpha_{eff}$ that is truly constrained by experiment. Equation~(\ref{alpefflab2}) thus implies that the actual coupling strength $\alpha$ can be much larger. With this revised perspective on the data, Fig.~\ref{expconsa}, for instance, says that laboratory experiments constrain $\alpha_{eff}\lsim 5\cdot10^{-4}$ (from Irvine~\cite{irvine}), implying the much weaker constraint on the bare coupling $\alpha\lsim 10^2$ for $\xi\sim {\cal O}(1)$. In particular, we should stress that $\alpha \sim {\cal O}(1)$, the expected outcome from string compactifications, is allowed.

\subsection{Planetary, Lunar and Geophysical Constraints}
\label{LargeScale}

The existence of a fifth force is also constrained by solar system data, in particular from the LAGEOS satellite, Lunar Laser Ranging (LLR), and anomalous perihelion precession of planets. See Fig.~\ref{expconsa}. For all of these, the effective interaction range is determined by the ambient baryonic gas and dark matter, with approximate density $\rho_G\approx 10^{-24}\;{\rm g}/{\rm cm}^3$. Thus Eq.~(\ref{lambdaeff}) implies 
\begin{equation}
\lambda_{eff} < 10\;(\xi\beta^2)^{-1/6}\;{\rm km}\,.
\end{equation}
Therefore, even for $\xi\sim 10^{-6}$ (and reasonable $\beta$), this gives $\lambda_{eff}\lsim 100$~km. From Fig.~\ref{expconsa}, on the other hand, we see that LLR and planetary data only significantly constrain a fifth force for ranges roughly greater than 1000~km. Hence, unless $\xi$ is exceedingly small, the constraints from LLR and planetary orbits become completely irrelevant.

One is therefore left with the LAGEOS data, a satellite orbiting the Earth at an average distance of approximately $10^4$~km. The radius of the Earth is of order $10^{10}$~mm, while that of the satellite is $\sim 10^3$~mm, resulting in an effective coupling strength of 
\begin{equation}
\alpha_{eff}\sim 10^{-28}\left(\frac{\alpha}{\xi^2}\right)^{1/3}\,.
\label{LAGEOS}
\end{equation}
We infer from Fig.~\ref{expconsa} that the LAGEOS constrains $\alpha_{eff}\lsim 10^{-4}$ for $\lambda_{eff}\lsim 100$~km. The above thus implies that this bound is consistent with a bare coupling constant as large as $\alpha\sim 10^{72}\xi^2$!

Geophysical constraints result from gravity experiments performed in mines, boreholes, lakes, oceans and towers. All of these essentially measure the relative difference in the gravitational force on two test masses at different depth. Since this involves the gravitational field of the Earth, substituting $R_{\oplus}\sim 10^{10}$~mm in Eq.~(\ref{betaeffbis}) gives
\begin{equation}
\alpha_{eff} < 10^{-22}\left(\frac{\alpha}{\xi^2}\right)^{1/3}\,.
\label{EarthsField}
\end{equation}
Once again it is clear from Fig.~\ref{expconsa} that these constraints can thus be neglected for reasonable parameter values.

\subsection{Isospin triplet coupling}
\label{triplet}

In section~\ref{NATURALNESS}, we concluded that the difficulties in constructing a model that satisfied field theoretic notions of naturalness could be ameliorated if $\beta$ were equal and opposite for protons and neutrons (and zero for electrons).  Such a situation can be termed an isospin triplet coupling of the scalar to matter.  Let us briefly review how this model compares with our main focus, namely a universal value of $\beta$ for all particles.

Clearly, the effective screening length $m_{eff}^{-1}$ now depends not on the ambient density $\rho$, but rather on the difference $\rho_p - \rho_n$ of the density $\rho_p$ due to protons and $\rho_n$ due to neutrons.  The key result (\ref{friendly}) carries over with the replacement $\rho \to |\rho_p - \rho_n|$.  In Table~\ref{lengths} we list some representative values of $m_{eff}^{-1}$ and $\beta_{eff}$ for an object 1~cm in radius, assuming $\xi = 1/10$ and $\beta = 1$.
\begin{table*}[htb]
\small
\hbox to \hsize{\hfil\begin{tabular}{|c|c|c|c|c|}
\hline
\hspace{6pt}Material	& $\rho$ [${\rm g}/{\rm cm}^3$] & $|\rho_p - \rho_n| / \rho$ & 	$m_{eff}^{-1}$ & $\beta_{eff}/\beta$ for $R_c=1$ cm\\
\hline
\hspace{6pt}lithium	& 0.4				& 0.14			     & $0.3\;{\rm mm}$	& $8\cdot 10^{-3}$ \\
\hspace{6pt}aluminum	& 2.7 & 0.04 & $0.3\;{\rm mm}$  	& $8\cdot 10^{-3}$\\	         	
\hspace{6pt} copper	& 8.9 & 0.1 & $0.15\;{\rm mm}$ & $2\cdot 10^{-3}$ \\ 
\hspace{6pt} gold	& 19.3 & 0.2 & $0.1\;{\rm mm}$ & $9\cdot 10^{-4}$ \\
\hspace{6pt}dry air	& $1.3 \times 10^{-3}$ & $1.3 \times 10^{-3}$ & 1\;{\rm cm} & -- \\
\hspace{6pt} ${{\rm interstellar} \atop {\rm medium}}$ & $10^{-24}$ & 1 & 
  $10\;{\rm km}$ & --\\
\hline
\end{tabular}\hfil}
\caption{Effective interaction range and coupling in various materials for the isospin triplet model}
\label{lengths}
\end{table*}
Dry air has an excess of neutrons, due mostly to the fact that argon makes up about a percent of air and has a $10\%$ excess of neutrons over protons.  Humid air picks up extra protons from water molecules.  This results in $|\rho_p - \rho_n|/\rho = 0$ when the dew point is roughly $17\,^\circ {\rm C}$, corresponding to a relative humidity of approximately $80\%$ at $20\,^\circ {\rm C}$.

In light of the sub-millimeter screening lengths in the first several entries of Table~\ref{lengths}, laboratory experiments of the type described in section~\ref{LabSearches} result in constraints that are not much different from the case where $\beta$ is the same for all particles. Indeed, most test masses used in fifth force searches are made of materials such as copper, beryllium, iron etc., all of which have a proton overdensity of about 10\%
as seen from Table~\ref{lengths}. 
This translates into an increase of about $10^{2/3}\approx 4.6$ in the effective coupling $\beta_{eff}$, as seen from Eqs.~(\ref{mmin0}) and~(\ref{betaeff}).
All the scaling laws, {\it e.g.}, $\alpha_{eff}\sim (\alpha/\xi^2)^{1/3}$ as in Eq.~(\ref{alpefflab2}), still hold. In particular, $\alpha \sim 1$ is still allowed, provided that $\xi\gsim 10^2$. 
Meanwhile, the analysis of planetary, lunar and geophysical constraints of section~\ref{LargeScale} is only mildly altered since the Earth consists primarily of iron with proton overdensity of about 5\%. Thus, $\alpha_{eff}$ remains ridiculously small, as in Eqs.~(\ref{LAGEOS}) and~(\ref{EarthsField}).

\section{Summary and Discussion} \label{conclu}

Figure~\ref{expconsbrevised} shows the revised constraints on $\alpha$ and $\lambda$ for $\xi=1$. The key result is that $\alpha\sim {\cal O}(1)$, corresponding to a fifth force of gravitational strength, is allowed for all scales probed by experiments.
Our model essentially escapes the E$\ddot{{\rm o}}$t-Wash and Irvine bounds because the test masses used in these experiments are large enough to suffer from a thin-shell suppression.

Nevertheless, we wish to stress that a modest improvement in current laboratory set-ups could lead to an unambiguous detection of the chameleon. Indeed, consider a laboratory experiment to detect a fifth force between two test masses of radius $R$ in vacuum. We have in mind a set-up similar to the E$\ddot{{\rm o}}$t-Wash experiment, for instance. Assuming $\beta=1$ and $\xi\lsim 1$, then Eq.~(\ref{betaeffbis}) gives 
\begin{equation}
\alpha_{eff} \sim 10^{-2} R^{-4}[{\rm mm}]\,. 
\label{betaeffbis2} 
\end{equation}
The E$\ddot{{\rm o}}$t-Wash experiment used test masses of characteristic size $R \sim 1\;{\rm cm}$. This gives $\alpha_{eff}\sim 10^{-4}$ (see Eq.~(\ref{alpefflab2})), which is just under the radar of current experiments, as detailed in Sec.~\ref{LabSearches}. Repeating the same experiment with test masses of characteristic size $R\sim 1\;{\rm mm}$ would instead give 
\begin{equation}
\alpha_{eff} \sim  10^{-2}\,,
\end{equation}
which is well within the sensitivity range of the old E$\ddot{{\rm o}}$t-Wash experiment. 

Experiments such as Stanford~\cite{stan} and Colorado~\cite{col} have used test masses smaller than 1~mm, which therefore do not suffer from any thin-shell suppression. Indeed, comparing Figs.~\ref{expconsb} and~\ref{expconsbrevised}, we see that the corresponding curves are left unchanged by the addition of a quartic self-coupling.
In particular, Fig.~\ref{expconsbrevised} shows that the Colorado experiment provides the tightest bound on $\alpha$ on small scales. Modest improvements in its sensitivity would allow Colorado to probe the theoretically interesting range of $\alpha \lsim 1$.

Thus, by making the test masses a bit smaller or improving the sensitivity, one might detect the fifth force described here---provided the couplings $\beta$ are not significantly smaller than the natural value, namely unity. This detection would unambiguously be identified as resulting from a chameleon field, since the above fifth force would only appear for sufficiently small test masses. This could easily be verified explicitly, by performing the experiment with different size test masses and checking for the $R^{-4}$ dependence predicted above. Moreover, as shown in Fig.~\ref{comp}, the fifth force is steeper than $1/r^2$ at short distances. (If the force were $1/r^2$ at short distances, then $\beta^{(actual)}/\beta$ would be constant.) This should result in a torsion-pendulum signal that is distinguishable from the usual Yukawa fifth force.

Because it is possible in the near future to start probing the interesting range of parameters for our model, it would certainly be helpful to refine our calculation of the expected chameleon force and make it more tailored to a realistic experiment. For instance, test masses used in an E$\ddot{{\rm o}}$t-Wash-like experiment are not spherically symmetric, as assumed here, but instead cylindrical slabs. We leave for future work the calculation of the precise force and expected signal for such a set-up.

\section*{Acknowledgments}

We thank C.G.~Callan, J.C.~Long, P.J.E.~Peebles, S.L.~Sondhi, A.~Weltman, and especially S.A.~Hoedl for helpful discussions. 
The work of SSG was supported in part by the Department of Energy under
Grant No. DE-FG02-91ER406571, and by the Sloan Foundation.
The work of JK was supported in part by the Columbia University Academic Quality Fund and the Ohrstrom Foundation.

\end{document}